# AnuraSet: A dataset for benchmarking Neotropical anuran calls identification in passive acoustic monitoring

## Authors

Juan Sebastián Cañas[1]*, Maria Paula Toro-Gómez[1], Larissa Sayuri Moreira Sugai[2], Hernán Darío Benítez Restrepo[3], Jorge Rudas[1], Breyner Posso Bautista[1], Luís Felipe Toledo[4], Simone Dena[5], Adão Henrique Rosa Domingos[6], Franco Leandro de Souza[7], Selvino Neckel-Oliveira[8], Anderson da Rosa[8], Vítor Carvalho-Rocha[8], José Vinícius Bernardy[9], José Luiz Massao Moreira Sugai[9], Carolina Emília dos Santos[9], Rogério Pereira Bastos[9], Diego Llusia[10,11,12], Juan Sebastián Ulloa[1]

## Affiliations

1. Instituto de Investigación de Recursos Biológicos Alexander von Humboldt, Avenida Paseo Bolívar 16-20, Bogotá, Colombia; 2. K Lisa Yang Center for Conservation Bioacoustics, Cornell Lab of Ornithology, Cornell University, 159 Sapsucker woods road, 14850, Ithaca, New York, USA; 3. Pontificia Universidad Javeriana Seccional Cali, Calle 18 No 118-250, Cali, Valle del Cauca, Colombia.; 4. Labortório de História Natural de Anfíbios Brasileiros (LaHNAB), Universidade Estadual de Campinas, Campinas, SP, BR; 5. Museu de Diversidade Biológica (MDBio), Universidade Estadual de Campinas, Campinas, SP, BR; 6.Instituto de Pesquisa da Biodiversidade (IPBio), Reserva Betary, Iporanga, São Paulo, Brazil; 7. Universidade Federal de Mato Grosso do Sul, Instituto de Biociências, Campo Grande, MS, BR; 8. Departamento de Ecologia e Zoologia, Universidade Federal de Santa Catarina, Florianopolis, SC, BR; 9. Universidade Federal de Goiás: Goiania, GO, BR; 10. Terrestrial Ecology Group, Departamento de Ecología, Universidad Autónoma de Madrid, C/ Darwin, 2, Ciudad Universitaria de Cantoblanco, Facultad de Ciencias, Edificio de Biología, 28049 Madrid, Spain; 11. Centro de Investigación en Biodiversidad y Cambio Global (CIBC), Universidad Autónoma de Madrid. C/ Darwin 2, 28049 Madrid, Spain; 12. Laboratório de Herpetologia e Comportamento Animal, Departamento de Ecologia, Instituto de Ciências Biológicas, Universidade Federal de Goiás, Goiás, Brazil

*corresponding author(s): Juan Sebastián Cañas (jcanas@humboldt.org.co)

## Abstract

Global change is predicted to induce shifts in anuran acoustic behavior, which can be studied through passive acoustic monitoring (PAM). Understanding changes in calling behavior requires the identification of anuran species, which is challenging due to the particular characteristics of neotropical soundscapes. In this paper, we introduce a large-scale multi-species dataset of anuran amphibians calls recorded by PAM, that comprises 27 hours of expert annotations for 42 different species from two Brazilian biomes. We provide open access to the dataset, including the raw recordings, experimental setup code, and a benchmark with a baseline model of the fine-grained categorization problem. Additionally, we highlight the challenges of the dataset to encourage machine learning researchers to solve the problem of anuran call identification towards conservation policy. All our experiments and resources can be found on our GitHub repository https://github.com/soundclim/anuraset.

## Background & Summary

Global anthropogenic biodiversity loss is a major challenge of contemporary society[1]. With severe wildlife population declines and extinctions over the earth, monitoring and predicting species responses to global changes became an urgent task for conservation. Novel technologies now offer remote, non-invasive, and automated methods to survey and monitor biodiversity at unprecedented spatial and temporal scales[2]. For instance, passive acoustic monitoring (PAM) has been largely adopted in ecological research and is increasingly used in conservation applications[3]. Based on acoustic sensor networks, PAM enables us to remotely and automatically record the vocal activity of wild animals, increasing our ability to study biological communities. However, a critical bottleneck for the widespread use of this method is the need for automated techniques to retrieve biologically meaningful information in the huge time-series audio datasets collected by PAM. Manual inspection of these recordings is unattainable due to the human specialist workload when audio data collected reach the big data scale [4].

In the last decade, the three fundamental reasons for the success of machine learning (ML) techniques have been the advancement in high-computing hardware, novel algorithms, and the curation of high-quality datasets for standardized benchmark[5]. As a consequence, ML has emerged as a key solution and a general accelerator for multiple domains in which biodiversity monitoring programs, animal ecology, and global change research are not an exception[6,7]. Particularly, the growth of ML for ecological applications now depends on the variety, quality, and availability of public datasets that define ML tasks for determined contexts and problems[7,8]. Despite recent efforts to curate datasets for ecological research, available data remains taxonomically and geographically biased[9]. ML has opened up exciting possibilities for research in this area[10–14], but limitations in the diversity of existing datasets must be acknowledged. In the field of bioacoustics and PAM, datasets aimed at supporting acoustic identification have been developed for a limited number of taxonomic groups, mainly birds[15,16], mosquitoes[17], and mammals[8,18,19]. These datasets have also served as general benchmarks for detection and classification of the recorded individuals into species[20]. Altogether, the increasing number of curated datasets coming from bioacoustics research generates a unique opportunity to foster the culture of open data, open models, and benchmarks in conservation research[21]. PAM has special importance in applied conservation, where datasets may impact the robustness of biodiversity monitoring programs that support ecological[22] and policy-related[23] decision-making.

Amphibians are one of the most endangered vertebrate groups in the world, with more than 40% of the species endangered to extinction[24]. In the tropics, amphibian communities exhibit high diversity[25] and are more prone to extinction[26] compared to other regions. To monitor these communities, researchers can take advantage of PAM techniques which are a non-invasive data collection that allows incorporating information from both rare and cryptic species, as well as from common and abundant ones. Acoustic communication has a central role in the reproductive behavior of anurans[27]. During the breeding season, males call, for example, to attract females, defend territories, and deter competitors[28]. Thus, a wide range of research relies on the identification and quantification of these sounds, with an increasing number of applications. However, there is a lack of open datasets for this highly vocal group that can support the development of ML models for PAM research.

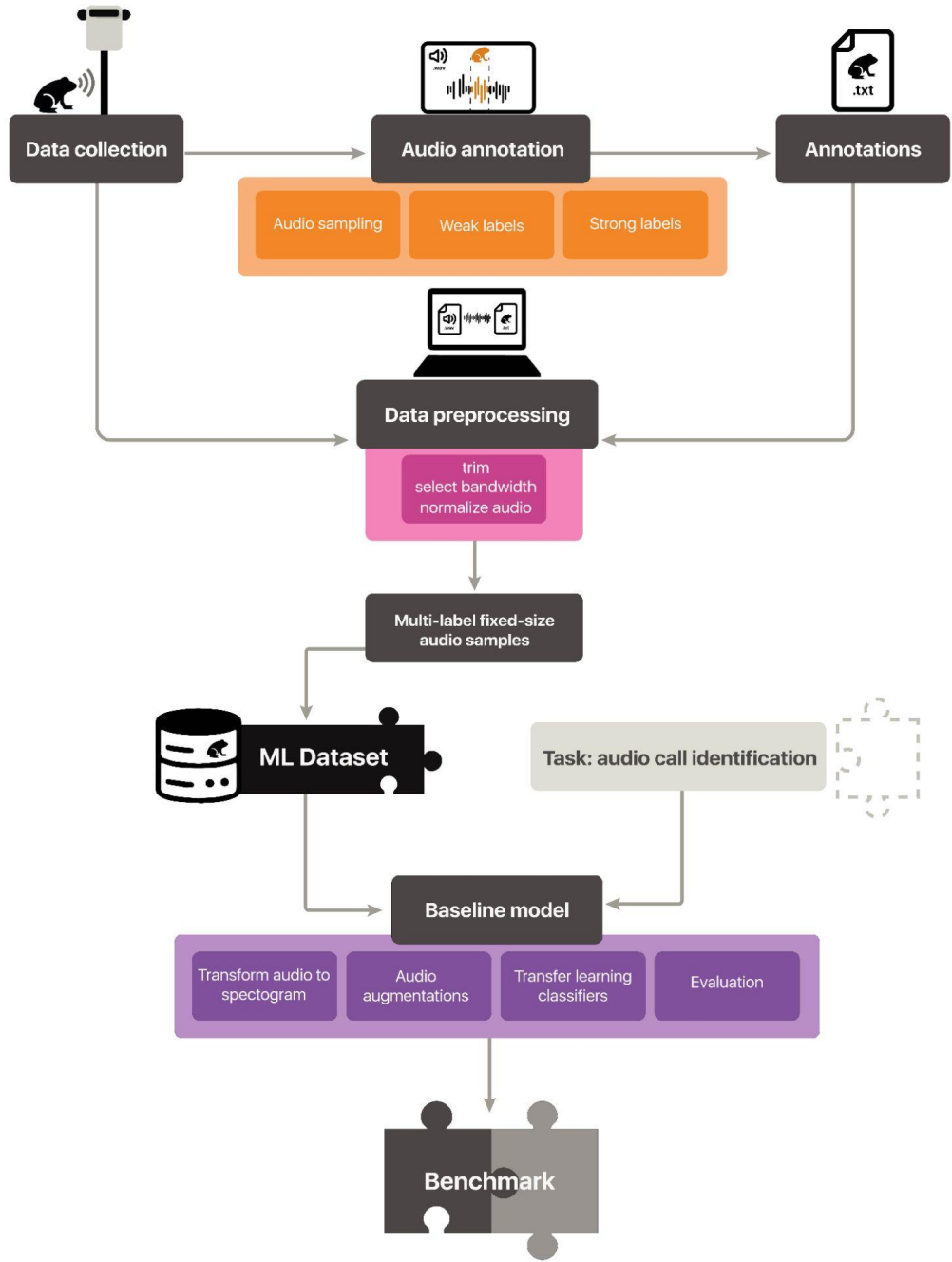

**Figure 1.** Overview of the AnuraSet methodological workflow that encompasses the process of dataset creation and benchmarking. It begins with the collection of passive acoustic monitoring data from four sites in the Neotropics. Subsequently, we annotated the acquired recordings with both weak and strong labels. Leveraging these annotations, we undertook a preprocessing of the data to construct a machine learning-compatible dataset. For solving the problem of anuran call identification, we frame the problem as a multilabel classification challenge, and to establish a baseline model, we adopted a transfer learning approach. Furthermore, we merged a specific task with the dataset, culminating in the creation of a benchmark.

This study introduces a large-scale annotated dataset of Neotropical anuran calls: AnuraSet. This dataset was compiled through a country-wide collaborative PAM program across Brazil between 2019 and 2021, and it is composed of 1612 1-minute annotated audio recordings, equivalent to 26.87 hours of audio. We collected data from four strategically selected sites in the Neotropics and collaborated with local scientists for precise annotation of the recordings. Subsequently, we preprocessed the data to train deep learning models, enabling us to conduct a baseline experiment and launch a benchmarking initiative for the automated identification of anuran calls (Figure 1). The preprocessing and baseline code is released under the MIT License and all the data is under the CC0 license to support reproducible research. AnuraSet will potentially provide a common and realistic-scale evaluation task for species identification in Neotropical soundscapes. In addition, AnuraSet is a solid starting point for a comprehensive and accessible dataset of anuran calls and choruses. Since tropical acoustic environments are highly complex and manually annotated datasets are scarce, AnuraSet has the capacity to accelerate the development of robust machine listening models for wildlife monitoring in biodiversity hotspots. Furthermore, we summarize the main challenges and propose a roadmap to foster a culture of collaboration, experimentation, research, and exploration in ML for applied ecology. In our viewpoint, this culture is essential for advancing ML techniques and ecological inferences for conservation policies. In addition, the challenges posed by biodiversity acoustic monitoring provide a unique opportunity for exploring new avenues in the field of ML.

In summary, our contributions are (i) a collection of manually annotated PAM recordings of Neotropical anurans calling activity, with information on species composition (presence-absence data) and audio quality of the recordings; (ii) a curated, preprocessed, and in the wild acoustic dataset, with a detailed description of the data challenges; and (iii) baseline models for benchmarking the problem of species identification towards the creation of robust classifiers and the fast development of new models. Overall, our goal is to support a community of ML researchers and conservationists who can work together to develop innovative solutions for biodiversity monitoring. By providing open-access resources and encouraging the exploration of new techniques, we aim to contribute to developing powerful tools for conservation and ecological research.

## Methods

### Data Collection

Calling activity of Neotropical anuran communities was monitored from 2019 to 2021 in four sites located at the Cerrado (INCT17, INCT41) and Atlantic Forest (INCT20955, INCT4) biomes, known for their critical role as global biodiversity hotspots (Figure 2). We call each station INCT to refer to *Institutos Nacionais de Ciência, Tecnologia e Inovação* (National Institutes of Science and Technology). At the edge of the water bodies of each site, we installed an acoustic sensor equipped with omnidirectional microphones (SM4, Wildlife Acoustics, Inc., Concord, MA, USA) that were fixed on trees or wooden bases, at about 1.5 m above the ground. Each recorder was configured to register one min every 15 min over 24h a day (a total of 3.2 hours per day), with a sampling rate of 22050 Hz and 16-bit depth resolution. Audios were also recorded in stereo mode, with 10 dB and 16 dB gain on each channel.

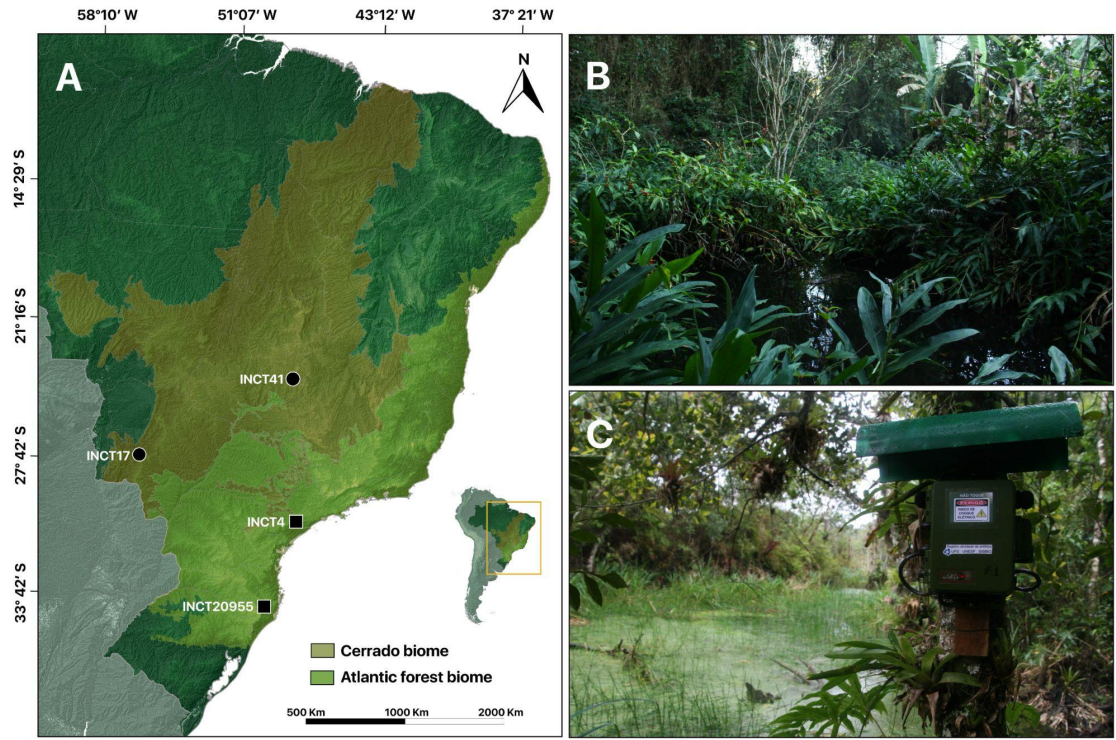

**Figure 2. A.** Geographic location of the four sites where the passive acoustic monitoring data were collected. Sites at the Cerrado biome INCT17 and INCT41 (dots) and at the Atlantic Forest biome INCT20955 and INCT4 (squares) **B.** Photograph of INCT4 monitoring site at The Atlantic forest biome. **C.** Acoustic sensor used to record anuran calls at the edge of a water body.

## Audio Annotation

We developed an annotation protocol in order to build automated tools for determining the species recorded with PAM. We combined weak labels (temporal precision was limited to the 60 s duration) with strong labels (providing exact temporal segments of the audio recording where the anuran call was active). The weak labels were annotated by local herpetologists and bioacoustics experts and the strong labels were annotated by a herpetologist over a selected subsample of all raw recordings to obtain a presence-absence dataset at the scale of the audio recording. All annotators had previous experience detecting anurans calls in recordings. Since the list of species at each study site was initially unknown, we first searched each 1-min recording of the species using local expert knowledge in the form of weak labels. After that, we used strong labels as they are better to solve the audio event classification problem[29]. The protocol that we developed consists of three steps specifically tailored for the identification of anuran calls. However, it can be easily adapted and customized for any taxon.

*Step 1. Audio sampling*

To annotate audio files, train, and validate ML models, we first obtained a random stratified sample of audio recordings from each site. This involved between 300 and 600 1-min audio files per site, selected among all the files collected during the local breeding season (3-6 months) at night time (from 1h before sunset until 1h before sunrise). To get a representative sample of recordings within the selected time frame, we divided the entire acoustic dataset into three seasonal periods (early, middle, and late season) and three daily periods (early, middle, and late night), and drew from 50 to 100 files per stratum. In total, we selected 1612 1-min audio files (26.87 hours) over the four study sites.

*Step 2. Weak labeling*

To identify anuran species recorded in the selected samples, local herpetologists and bioacoustics experts (JVB, SD, JLMMS, AdaR) performed a visual and auditory analysis of spectrograms using the Audacity ® 3.2.5 software[30], which was used in all steps of the annotation protocol. Local annotators were tasked to report the level of calling activity of each recorded anuran species, based on the Amphibian Calling Index[31] (Table 1), according to the detection of their advertisement call in each 1-minute audio file (weak labeling).

**Table 1**. Levels of anuran calling activity for the weak labeling.

| Calling activity level | Score | Description of chorus activity |
|---|---|---|
| Absence | 0 | No anuran calls recorded |
| Low | 1 | Individual anurans can be counted and calls do not overlap |
| Moderate | 2 | Few anuran individuals that cannot be accurately counted, and with both separate and overlapping calls |
| High | 3 | Intense chorus, with continuous and overlapping calls from different individuals |

*Step 3. Strong labeling*

To provide precise annotations, we identified bouts of advertisement calls within each audio file and generated strong labels for them (step 1). Using Audacity 3.2 software, we conducted a detailed visual and aural inspection of the spectrogram to identify temporal limits (beginning and end) of audio segments containing species-specific calls with an inter-call interval of less than 1 second. These annotations ensured fine-scale specificity (Figure 3). For longer intervals, we split the calls into different time boxes and labeled them independently. Detailed labels assigned to time boxes were composed of (i) the species ID, tagged with a unique 6-letter code built from the scientific name of each identified species (Table 2), and (ii) the perceived quality of the recorded signal, included as a single letter indicating a Low ('L'), Medium ('M'), or High ('H') quality (Figure 4). To ensure consistency among the perceptual quality labels, we set up the following criteria: A high-quality call has a high signal-to-noise ratio, no overlap with other sounds, has a well-identifiable structure on the spectrogram, and can be easily visualized on the oscillogram. A medium-quality call can be visually identified on the spectrogram but may overlap with other sounds that can be difficult to identify in the oscillogram. A low-quality call shows a low signal-to-noise ratio, is partially masked by other sounds, appears with low intensity on the spectrogram, and cannot be easily identified on the oscillogram. This information was used to increase the usability of the data and improve the error analysis of the learning model.

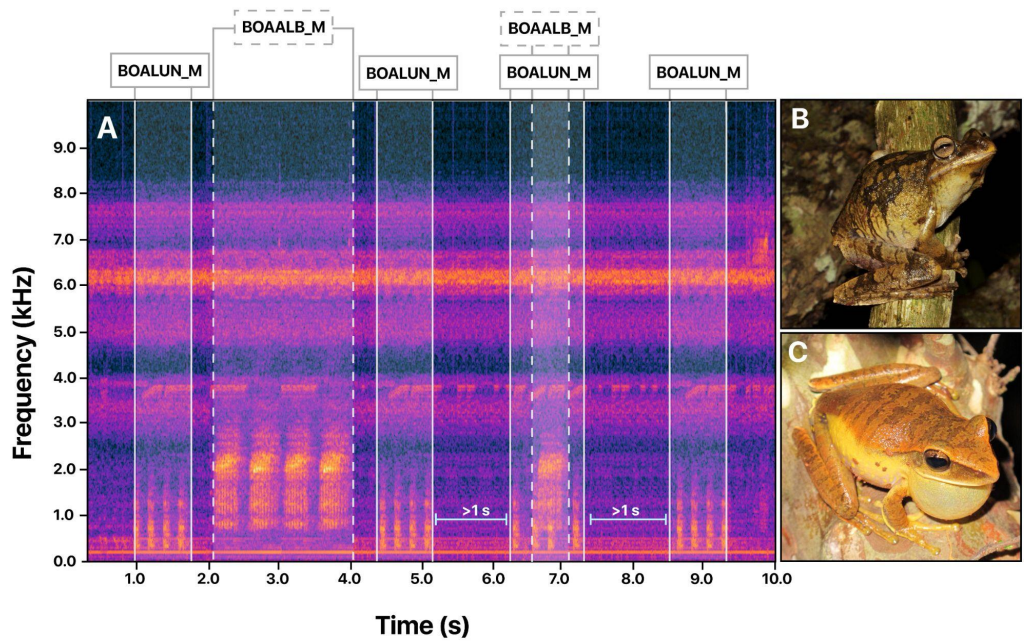

**Figure 3. A.** Example of strong labeling. For each 1-min raw audio file sampled, the herpetologist annotator identified and selected the temporal limits of the advertisement call. We annotated calls using different time selections when spaced more than 1 second apart. **B.** Image of an individual of *Boana lundii* its advertisement call coded as BOALUN in the annotation process **C.** A calling male of *Boana albopunctata* (BOAALB).

We followed a consistent annotation procedure for all the data, performed by a single trained herpetologist (MPTG). We used Audacity ® 3.2.5 software[30] to visualize the spectrograms and create the labels in steps 2 and 3. We optimized the visualization of the acoustic signals by setting the spectrogram configuration parameters as follows: linear scale for frequency, the maximum frequency of 10 kHz, gain of 20 dB, range 80 dB, FFT algorithm with a window size of 1024, and standard color range to represent sound energy.

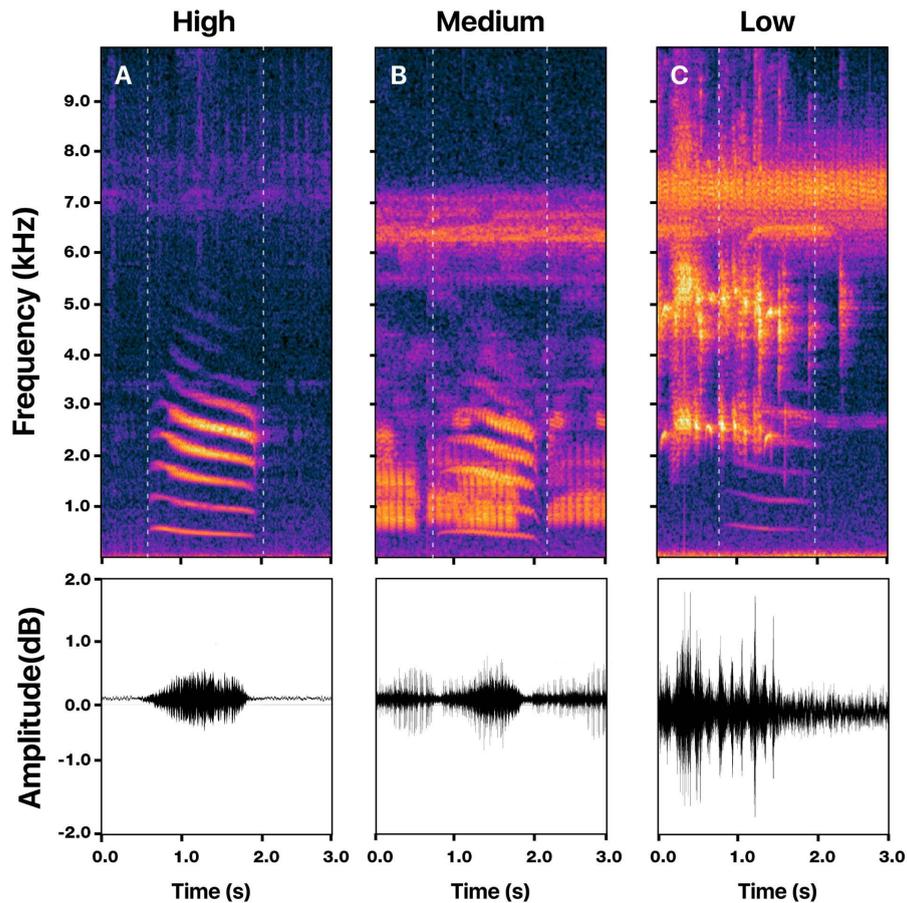

**Figure 4.** An illustrative example of the advertisement call of *Physalaemus albonotatus* for the three audio quality categories. **A.** High-quality call ('H') shows a high signal-to-noise ratio, no overlap with other sounds, has a well-identifiable structure on the spectrogram, and can be easily visualized on the oscillogram. **B.** Medium-quality call (M) can be visually identified on the spectrogram but may overlap with other sounds that can be difficult to identify in the oscillogram. **C.** Low-quality call (L) shows a low signal-to-noise ratio, is partially masked by other sounds, appears with low intensity on the spectrogram, and cannot be easily identified on the oscillogram.

### Data Preprocessing

We framed the species identification problem as a multi-label classification task considering the common occurrence of call overlap in PAM. We applied a set of transformations over the raw audio files and annotations to obtain a dataset suitable for use with ML algorithms. First, reading the metadata of the 1-minute raw audio files we obtained samples of a 3-second fixed-length window applying a 1-second sliding window. This produces a 75% overlap between samples[10,32]. Second, we assigned a multilabel species label to each sample if any portion of a species call appeared within one of these windows, we considered it as an occurrence for that species, and this procedure was applied to all calls, regardless of their quality. Third, we preprocessed each 1-minute annotated audio file using the scikit-maad python package[33] applying the sliding window approach described above. After trimming the 3-second in time and the frequency limits between 1Hz and 10000Hz, we applied a bandwidth filter which uses a bandpass filter to process a 1d signal with an infinite impulse response (IIR) Butterworth filter of order 5; after that, we normalized the audio signal to a maximum amplitude of 0.7 decibel full-scale value (dBFS) and saved as uncompressed WAV format. Finally, we selected each 1-minute recording with weak labels to split the dataset between training and test, summed all the species occurrences, and applied an iterative stratification for the multi-label setting[34,35] to the unbalanced proportions in the different subsets, with 70% in training and 30% in the test. In this step, we used the 1-minute recording level to avoid data leakage (same 1-minute audio with samples in train and testing subsets).

## Data Records

We collected data for 42 neotropical anuran amphibian species from 12 genera and 5 families (Table 2). Taxonomic nomenclature followed Frost[36]. The annotations for all individual or series of calls from these species are 16,075 time boxes equivalent to approximately 31 hours of cumulative duration and 27 hours of human-generated annotations. It is important to note that due to significant overlap in time boxes among different species, the cumulative duration exceeded the sum of the recording time. Among the collected data, approximately 20% of the 1-minute raw audio files did not contain anuran calls but a soundscape from geophonic sources like rain and wind, as well as biophonic sources such as other vocalizing species like insects and birds. The annotated data was distributed across the sites INCT17, INCT20955, INCT41, and INCT4 as 42.5%, 33%, 13.5%, and 11%, respectively. The distribution of samples per species in the final dataset exhibits a long-tailed pattern, which coincides with the typical species diversity pattern in tropical environments (Figure 5). Additionally, we observed a high degree of variability in species composition between sites; specifically, only five species were detected in more than one site.

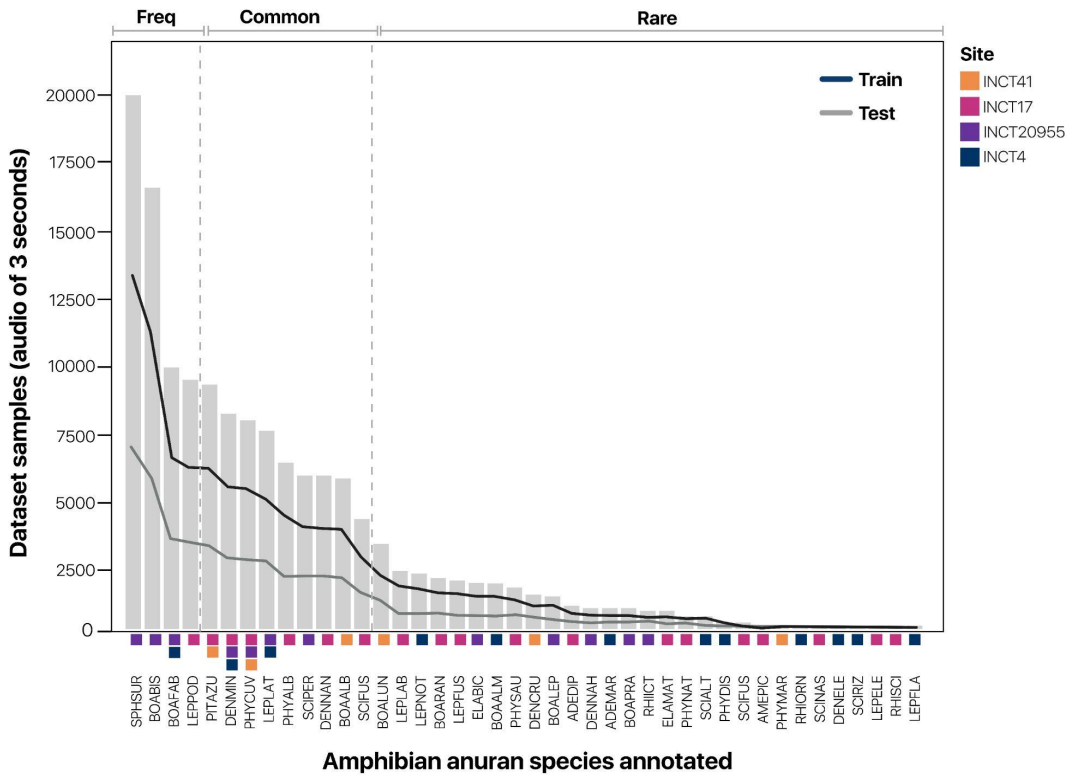

**Figure 5.** Frequency distribution of 3-second samples per anuran species. The long-tailed distribution is a typical distribution of a real-world species diversity dataset. We split species into the classes of 'common', 'frequent', and 'rare' to determine the effect of sample size on the performance of the species identification problem. Additionally, the occurrence of the same species in different sites is represented by different colored squares at the bottom of the histogram. The training and test set distributions obtained by using the split strategy are depicted with black and blue lines, respectively.

Here we provide two main data resources: (i) the raw audio files with an associated table containing annotations, and (ii) the preprocessed input dataset for ML with 93378 3-second audio samples, both sharing a similar folder structure. The raw data was divided into separate folders per site. Inside each folder, there is a collection of 1-minute recordings in WAV format with self-explanatory filenames that include the site name, the date, and the time as follows: **{site}_{date}_{time}.wav**. For example, the file INCT20955_20190830_231500.wav is located in the folder of site INCT20955 and was obtained on 30 August 2019 at 23:15 (BRT time zone). In the same way, the preprocessed dataset follows the same folder and naming structure but also includes the start and final second of the audio segment: **{site}_{date}_{time}_{start second}_{final second}.wav.** Following the previous example, INCT20955_20190830_231500_30_33.wav means that the sample starts in the second 30 and ends at the second 33. The dataset folder contains 2 files and one folder containing separate folders per site. The samples are WAV audio files with fixed 3-second lengths, obtained with 22.05 kHz sampling frequency and 16-bit depth. The two other files are a README file describing the structure and construction of the dataset and a metadata CSV file containing the labels for each sample as follows:

- sample_name: the unique identifier of each sample that corresponds to a unique audio file in the audio folder and follows the structure **{site}_{date}_{hour}_{start second}_{final second}.wav.** The next 5 columns were constructed based on this column.
- fname: raw audio filename extracted from a site and used by annotators to create weak labels.
- min_t: second where the annotation starts in a fixed window length.
- max_t: second where the annotation ends in a fixed window length.
- site: identifier of the recording site.
- date: datetime of the recording.
- subset: training or test subset.
- species_number: total number of species in each sample. The sum of the next 42 columns per row.
- **{species}**×42 Binary columns of each species where 1 if some portion of the call is in the sample, 0 else. The 42 species column names are the codes shown in Table 2.

The dataset and the raw data are provided under the Public Domain Dedication license (CC0) and are hosted in Zenodo with the DOI 10.5281/zenodo.8056090.

**Table 2.** List of the 42 anuran species included in the AnuraSet. Species code was built from the scientific name of each identified species, following the first three letters of the genus, followed by the first three letters of the specific epithet. Hyperlinks provide access to species pages in the Global Biodiversity Information Facility.

| SPECIES NAME | SPECIES CODE |
|---|---|
| **BUFONIDAE** | |
| *Rhinella icterica* | RHIICT |
| *Rhinella ornata* | RHIORN |
| *Rhinella scitula* | RHISCI |
| **DENDROBATIDAE** | |
| *Ameerega picta* | AMEPIC |
| **HYLIDAE** | |
| *Boana albopunctata* | BOAALB |
| *Boana albomarginata* | BOAALM |
| *Boana bischoffi* | BOABIS |
| *Boana faber* | BOAFAB |
| *Boana leptolineata* | BOALEP |
| *Boana lundii* | BOALUN |
| *Boana prasina* | BOAPRA |
| *Boana raniceps* | BOARAN |
| *Dendropsophus cruzi* | DENCRU |
| *Dendropsophus elegans* | DENELE |
| *Dendropsophus minutus* | DENMIN |
| *Dendropsophus nahdereri* | DENNAH |
| *Dendropsophus nanus* | DENNAN |
| *Pithecopus azureus* | PITAZU |
| *Phyllomedusa distincta* | PHYDIS |
| *Phyllomedusa sauvagii* | PHYSAU |
| *Scinax alter* | SCIALT |
| *Scinax fuscomarginatus* | SCIFUS |
| *Scinax fuscovarius* | SCIFUV |
| *Scinax nasicus* | SCINAS |
| *Scinax perereca* | SCIPER |
| *Scinax rizibilis* | SCIRIZ |
| *Sphaenorhynchus surdus* | SPHSUR |
| **LEPTODACTYLIDAE** | |
| *Adenomera diptyx* | ADEDIP |
| *Adenomera marmorata* | ADEMAR |
| *Leptodactylus elenae* | LEPELE |
| *Leptodactylus flavopictus* | LEPFLA |
| *Leptodactylus fuscus* | LEPFUS |
| *Leptodactylus labyrinthicus* | LEPLAB |
| *Leptodactylus latrans* | LEPLAT |
| *Leptodactylus notoaktites* | LEPNOT |
| *Leptodactylus podicipinus* | LEPPOD |
| *Physalaemus albonotatus* | PHYALB |
| *Physalaemus cuvieri* | PHYCUV |
| *Physalaemus marmoratus* | PHYMAR |
| *Physalaemus nattereri* | PHYNAT |
| **MICROHYLIDAE** | |
| *Elachistocleis bicolor* | ELABIC |
| *Elachistocleis matogrosso* | ELAMAT |

## Technical Validation

### Experimental setup

The main goal for creating the AnuraSet is to provide a solution for the 'species identification problem' and boost ecological inferences in PAM-based anuran monitoring programs. We frame the 'species identification problem' as a multi-label classification problem using the data from all 4 sites without temporal or site distinction. Following the ecological conditions of the large-scale analysis bioacoustics project[32], we choose the F-measure as the performance classification metric. For the case of multi-label classification, we selected the Macro version of the F-1 score to give the same importance to all species. To better understand the dependency between the number of samples and performance, we grouped species into 'Common', 'frequent', and 'rare' categories using the samples frequency similar to the Auto Arborist Dataset[12]. The grouping reflected the label frequency within each anuran assemblage with 2 breakdowns, where before 10.000 samples were classified as common species, less than 5.000 samples were classified as rare species, and those between 5.000 and 10.000 samples were classified as frequent species.

### Baseline Models

Following the pipelines of previous studies[10,32], we applied a Mel Spectrogram transformation on audio recordings using a window size of 512, a hop length of 28, and the number of mel filter banks of 128. Then we applied SpecAugmentation in time and frequency[37] as spectrogram augmentation strategies and resize. The transformations and augmentations described above generated the inputs in ResNet[38] family models. Specifically, we tested the ResNet18, ResNet50, and ResNet152. All our baseline experiments were implemented using the PyTorch[39] framework and the torchaudio[40] library which are publicly available in the repository https://github.com/soundclim/anuraset.

### Benchmark Results

**Table 3**. Performance of ResNet family models in F1 Score percentage using all sites and species.

|  | Macro F1-Score (%) ↑ | | |
| --- | --- | --- | --- |
|  | ResNet18 | ResNet50 | ResNet152 |
| Frequent | 61.6 | 62.3 | **68.4** |
| Common | 52.0 | 53.9 | **56.8** |
| Rare | 14.8 | 9.9 | **15.7** |
| All | 34.9 | 33.2 | **37.8** |

After testing the ResNet family models, we grouped the performance by species according to their classes of sample frequency (Table 3). The best model in all cases was the ResNet152, with a percentage (%) F1-score of 68.4 for the Frequent group, 56.8 for the Common, and 15.7 for the Rare classes. The total Macro F1-score was 37.8. This result suggested that the number of samples strongly influences the general performance of the models. The F1-score performance of each species in each site is reported in Figure 6. In this Figure, we confirmed the challenge of *learning from small samples*, which is related to the problem of creating machine learning models using just a few samples for training, for example in Figure 6 we can see that in less than 1000 samples the algorithms perform percentage F1- score less than 20% in all cases. This problem is still an open research area in deep learning for computational bioacoustics[41].

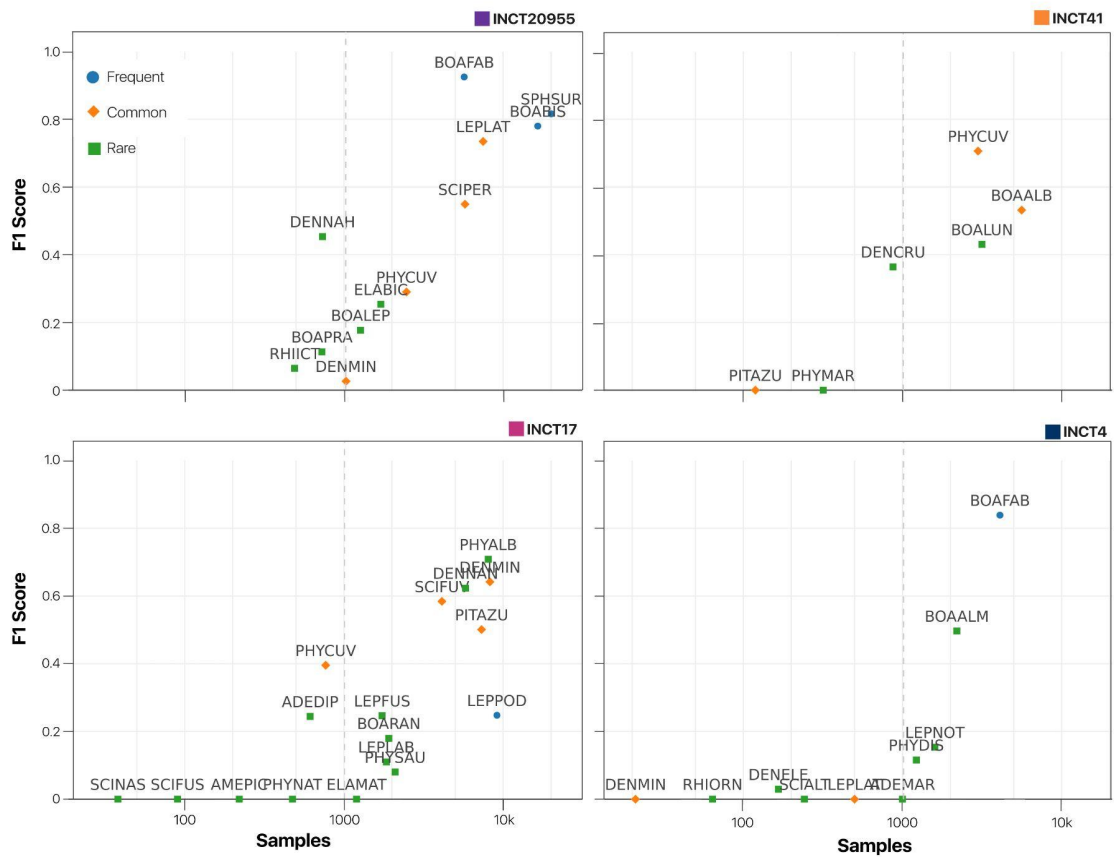

**Figure 6**. Performance for benchmarking the species identification problem. Using the ResNet152 model, we evaluated the species identification problem (see section 'Experimental Setup') in each site. The x-axis is the number of samples in the logarithmic scale and the y-axis is the F1 score. Across sites, we found a positive relationship between samples and performance.

## Usage Notes

### Data Challenges and Open Problems

During the annotation and dataset-building process, we faced challenges inherent to Neotropical, real-world datasets in PAM. We encourage researchers to experiment with the AnuraSet, from heuristics to understand the optimal parameters in preprocessing steps, including augmentation strategies to novel techniques for advancing the anuran call identification problem and other tasks yet to be discovered. With the goal of paving the way for new directions and advancements in ML research for bioacoustics and ecoacoustics, we summarize these challenges in the following topics.

**The devil is in the tails[42]**

As expected, the number of audio samples per species is highly imbalanced (Figure 5), forming a long-tailed distribution. The characteristics of a large number of categories and small training examples pose a challenge for obtaining good classifiers in all species. As we see in the benchmark results, there was a dependency between samples and performance. This situation is especially relevant when rare species are of interest for ecological and conservation applications. AnuraSet is a suitable dataset to test different methods such as algorithmic solutions[43–45] or augmentation strategies that have been proposed to overcome the long-tailed problem. Furthermore, this problem can be formulated as a *Learning from small samples* problem to explore state-of-the-art approaches[41] like few-shot learners[46,47] or self-supervised learning[48,49].

**Human Intensive Annotation**

Another manifestation of the *Learning from small samples* challenges happens in the early beginning of the annotation process. As we showed in the annotation protocol section, this is a human labor-demanding process. To scale in rich and large datasets it is necessary to use new ways to annotate data points as have been shown in clever approaches like Auto Arborist Dataset[12]. For example, one possible path is to explore a hybrid approach for human-machine collaboration labeling using an active labeling and learning scheme where each step of the learning procedure is actively assisted by a learning algorithm[50]. Recent work[51] shows that weak labels combined with unsupervised learning approaches can improve the performance of classifiers. The evaluation of such methods on the AnuraSet dataset can facilitate advancements in efficient and scalable annotation techniques.

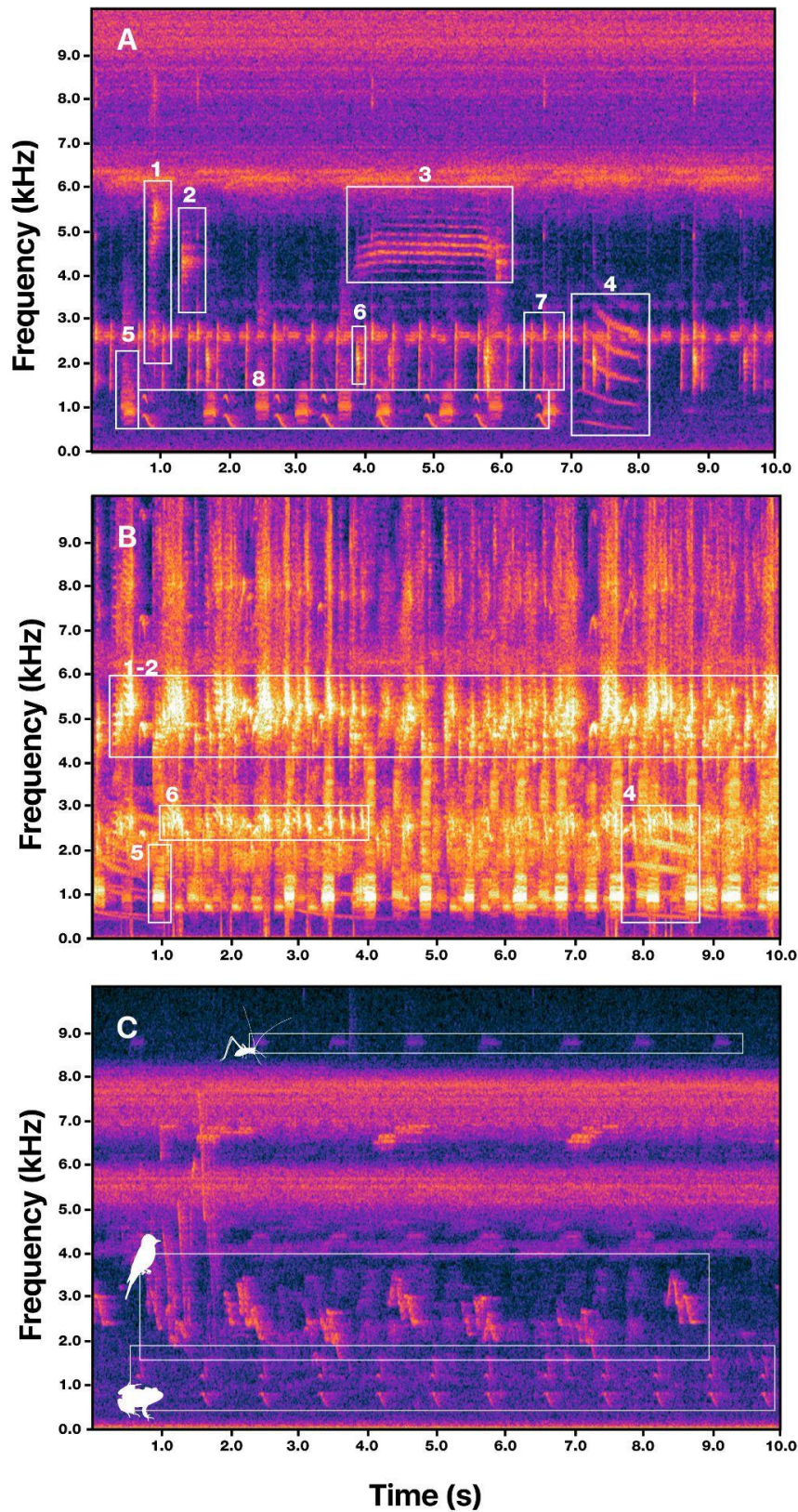

**Figure 7**. Analytical challenges of the AnuraSet: **A.** spectrogram showing eight different species that were recorded calling in less than eight seconds, highlighting the degree of co-occurrences, call overlap, and fine-grained identification; **B.** spectrogram showing a dense chorus with the low-signal-to-noise ratio and highly sound masking. **C.** spectrogram showing the richness and co-occurrence of sounds from different taxa (silhouettes from bottom to top depicting frogs, birds and orthopterans, respectively).

**Fine-grained audio in natural environments**

Despite a classic dataset such as ImageNet[52], where the classes can be easily identified for a human, the classes annotated in the AnuraSet rely on the expert knowledge of local herpetologists on sound-based species identification. Additionally, the recordings were collected in complex environments, generating variability in the signal-to-noise ratio of the data due to neotropical soundscape diversity in the different biomes (Figure 7). We confirm that the presence of calls in noisy conditions is a typical situation encountered in tropical environments investigated by PAM. This kind of problem, which involves distinguishing between subtle differences may imply other approaches[53,54], compared with generic object recognition.

**Multi-label dataset**

Tropical anuran assemblages recorded via PAM exhibit a distinctive feature of dense choruses with high call overlap, comprising different call types. This characteristic often leads to sound masking and makes the identification of individual calls challenging. Species calls in PAM recordings from AnuraSet are highly overlapped, therefore, calls often overlap not only between conspecifics but also between heterospecifics. As Figure 7A showed, 8 different anuran calls were recorded in less than 8 seconds. This characteristic is unique to PAM data and poses a challenge that is different from other wildlife monitoring sensors like camera trap images. These overlaps are related to the classic problem of the *cocktail party,* in which we try to search for an audio signal of interest like the anuran call, while other species, geophony, and biophony sound co-occur or overlap with the signal of interest. Recent studies[55–57] show promising progress in the context of bioacoustics.

**Towards abundance and behavior classification**

In the weak labeling process, we go beyond binary presence-absence annotation and use four categories to capture calling activity, similar to the Amphibian Calling Index[31] (Table 1). By mixing this assignment of weak labels with strong labels in the AnuraSet it is possible to work towards call activity classifiers that can measure anuran abundance and behavior in an ecologically meaningful way. These classifiers could help us understand species co-occurrence, temporal patterns of vocal activity, and chorus formation. Measuring abundance in bioacoustics is not straightforward, as it depends on factors such as variability of animal vocalization behavior, overlap, and interference of sounds from different sources. However, the AnuraSet provides a dataset with these properties in a natural and complex environment that will allow the development of new classification techniques that consider sources of error and bias.

## Code Availability

The dataset and the raw data are hosted in Zenodo https://doi.org/10.5281/zenodo.8056090 under the CC0 license. All the code for reproducing the experimental protocol, the building and preprocessing of the dataset, and the use of the baseline model are available in the repository https://github.com/soundclim/anuraset under the MIT license. We open the Python code to fast development of new deep learning models and experiments in Pytorch.

## Acknowledgements

The authors acknowledge financial support from the intergovernmental Group on Earth Observations (GEO) and Microsoft, under the GEO-Microsoft Planetary Computer Programme (October 2021); São Paulo Research Foundation (FAPESP #2016/25358-3; #2019/18335-5); the National Council for Scientific and Technological Development (CNPq #302834/2020-6; #312338/2021-0, #307599/2021-3); National Institutes for Science and Technology (INCT) in Ecology, Evolution, and Biodiversity Conservation, supported by MCTIC/CNpq (proc. 465610/2014-5), FAPEG (proc. 201810267000023); CNPQ/MCTI/CONFAP-FAPS/PELD No 21/2020 (FAPESC 2021TR386); Comunidad de Madrid (2020-T1/AMB-20636, Atracción de Talento Investigador, Spain) and research projects funded by the European Commission (EAVESTROP–661408, Global Marie S. Curie fellowship, program H2020, EU); and the Ministerio de Economía, Industria y Competitividad (CGL2017-88764-R, MINECO/AEI/FEDER, Spain).

## Author contributions

JSC and JSU conceived and designed the experiments; JSU, LSMS, and DL directed the project JSU, LSMS, DL, HB, RPB, and LFT obtained funding sources; RPB, JVB, LFT, SD, FLS, SNO, AR, VCR, CES, and AHRD participated in the data collection; MPT, JSU, LSMS, and DL designed the annotation protocol; JVB, SD, JLMMS, and AR made the weak labels, MPT made the strong labels; JSC, JSU, HB, JR, and BP participated in the benchmark and dataset design; JSC developed the code for dataset building, preprocessing, benchmark, and analysis tools; JSC wrote the original draft; JSC, MPT, JSU, LSMS, DL, and HB wrote and edited the draft; all authors reviewed the final draft.

## Competing interests

The authors declare no competing interests.